# NEAR-FIELD SAR IMAGE RESTORATION BASED ON TWO DIMENSIONAL SPATIAL-VARIANT DECONVOLUTION

*Wensi Zhang, Xiaoling Zhang, Xu Zhan, Yuetonghui Xu, Jun Shi, Shunjun Wei*

School of Information and Communication Engineering
University of Electronic Science and Technology of China,
Chengdu, China 611731

## ABSTRACT

Images of near-field SAR contains spatial-variant sidelobes and clutter, subduing the image quality. Current image restoration methods are only suitable for small observation angle, due to their assumption of 2D spatial-invariant degradation operation. This limits its potential for large-scale objects imaging, like the aircraft. To ease this restriction, in this work an image restoration method based on the 2D spatial-variant deconvolution is proposed. First, the image degradation is seen as a complex convolution process with 2D spatial-variant operations. Then, to restore the image, the process of deconvolution is performed by cyclic coordinate descent algorithm. Experiments on simulation and measured data validate the effectiveness and superiority of the proposed method. Compared with current methods, higher precision estimation of the targets' amplitude and position is obtained.

## 1. INTRODUCTION

Near-field synthetic aperture radar (SAR) has been widely used in the automatic target recognition, concealed weapon detection and radar-terrain interaction fields [1]-[3]. However, a large number of sidelobes and background clutter around the target exist in the SAR image, causing image quality degradation. Image restoration is necessary for better image comprehension.

Current SAR image restoration methods can be roughly divided into the following three types. The first one is the sparse reconstruction method, such as Iterative Shrinkage Threshold Algorithm (ISTA) [4]. However, in this kind of method, the restoration process is considered as a denoising process, which omits the features of sidelobes. Thus, when strong targets coexist with weak targets, the latter are easily recognized as noise.

The second type is spectral estimation method, such as Iterative Adaptive Approach (IAA) [5]. The degradation operation is considered under spatial-invariant approximation. This approximation is not applicable for the near-field SAR image. In near-field SAR, the restoration operation is spatial-variant both in range and azimuth direction. And the model mismatch causes residual clutters.

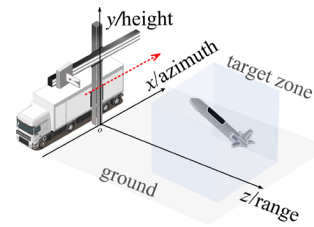

Fig.1 Near-field SAR imaging geometry.

The last type method is greedy method, also known as relax method, such as the CLEAN method [6]. With spatial-variant operation into consideration, its performance exceeds others. However, when targets are closely spaced, amplitude estimation bias is inevitable.

To obtain the image with higher quality, where the targets can be restored accurately with hardly any clutter remains, we proposed a near-field SAR image restoration method based on 2D spatial-variant deconvolution. The degraded image is considered as a complex convolution process with 2D spatial-variant operations. Through least square estimation by cyclic coordinate descent algorithm, the targets are deconvoluted.

Simulated and measured data are used for experimental validation. Results show that all the degraded images are well restored. Compared with three other methods (ISTA, IAA and CLEAN), the superior performance is achieved by the proposed method. All the targets' amplitudes and positions are restored accurately, even when they are closely spaced or when weak targets exist.

## 2. IMAGE DEGRADATION PROCESS

Figure 1 shows the geometry of near-field SAR imaging. The antenna is fixed on the scanning rail, and moves along azimuth direction. The target's spatial spectrum is received during the imaging process. However, due to limited length of the rail, only the partial spectrum is received, causing the degradation. Following the Born approximation, the target is a combination of multiple isolated scattering centers. Because of degradation, the scattering centers end up with limited

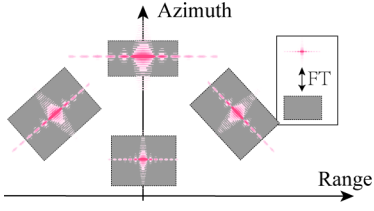

Fig.2 Spatial-variant degradation in near-field SAR.

resolution and surrounded by sidelobes. So, the image degradation model can be expressed as (1).

$$\mathbf{Y} = \sum_{i=1}^{M} \mathbf{D}_i * \mathbf{X}_i + \mathbf{N}$$
$$\mathbf{X} = \sum_{i=1}^{M} \mathbf{X}_i \quad (1)$$

where $\mathbf{Y}$ is the degraded image and $\mathbf{X}$ is the non-degraded image. $\mathbf{X}_i$ is the non-degraded image for a single point-like target, $\mathbf{D}_i$ is the corresponding degradation operation, $\mathbf{N}$ is the clutter and noise.

The degradation operation is related the received spatial spectrum of the target, which can be denoted as:

$$B_r = B_T, \quad B_a = \frac{\lambda R}{2L} \quad (2)$$

where $B_r$ is the range bandwidth that equals transmitting signal bandwidth $B_T$. And $B_a$ is the azimuth bandwidth. The limited rectangular shape spectrum degrades a single scatter from the ideal point-like pattern to 2D sinc-like. The degradation operation is obtained by the inverse Fourier Transform of the spectrum [7].

Further, as shown in Fig.2, the spectrum's direction is related to the observation angle. The range sidelobe follows its observation direction, while the azimuth one is vertical to it. And the variation of it, different from the far-field SAR, can't be ignored in the near-field SAR, especially when the imaging observation angle is large. Besides, seen from (2), the azimuth bandwidth is related to the target's slant range, which causes that the azimuth resolution is also spatial-variant.

To sum up, different from far-field SAR image, in near field, image degradation presents unique features. Because of 2D spatial-variant degradation operations, the resolution of target is varied along the range and sidelobes of different orientations coexist that may even bury the targets. Besides, clutter from the ground is unavoidable. Without the further restoration, they hinder the application of near-field SAR.

## 3. IMAGE RESTORATION PROCESS

As analyzed in the last section, the degraded image is the complex convolution of 2D spatial-variant degradation operations. Thus, through deconvolution, the image can be restored, which can be formed as:

$$\operatorname{argmin}_{\mathbf{X}} \frac{1}{2} \left\| \mathbf{Y} - \sum_{i=1}^{M} \mathbf{D}_i * \mathbf{X}_i \right\|_{\mathbf{F}}^2 \quad s.t. \quad \mathbf{X} = \sum_{i=1}^{M} \mathbf{X}_i \quad (3)$$

where $\|\cdot\|_{\mathbf{F}}^2$ denotes the Frobenius norm. And by defining $\mathbf{x}_s = \begin{bmatrix} \mathbf{x}_1^H & \cdots & \mathbf{x}_M^H \end{bmatrix}^H$, a stack of vectorized targets' images, and $\mathbf{D}_s = \begin{bmatrix} \mathbf{D}_1' & \cdots & \mathbf{D}_M' \end{bmatrix}$, a stack of convolutional operations, where $\mathbf{D}_i'$ is the matrix form of the convolutional operation. that satisfies $\mathbf{D}_i' \mathbf{x}_i = \operatorname{vec}(\mathbf{D}_i * \mathbf{X}_i)$, we can reformulate (3) as:

$$\operatorname{argmin}_{\mathbf{x}} \frac{1}{2} \|\mathbf{y} - \mathbf{D}_s \mathbf{x}_s\|^2 \quad s.t. \quad \mathbf{x} = \sum_{i=1}^{M} \mathbf{x}_i \quad (4)$$

where $\|\cdot\|$ denotes the $L_2$ norm, and $y$ is the vectorized degraded image.

As we know, the targets are actually sparse in the SAR image [8]. So we can reformulate (4) as (5) by adding a $L_1$ regularization.

$$\operatorname{argmin}_{\mathbf{x}} \frac{1}{2} \|\mathbf{y} - \mathbf{D}_s \mathbf{x}_s\|^2 + \lambda \|\mathbf{x}\|_1 \quad s.t. \quad \mathbf{x} = \sum_{i=1}^{M} \mathbf{x}_i \quad (5)$$

where $\|\cdot\|_1$ denotes the $L_1$ norm, and $\lambda$ is the regularization coefficient.

We adopt the cyclic coordinate descent algorithm [9] to solve it. Each target is repeatedly deconvoluted. In the $p^{th}$ iteration, we update the target $\mathbf{x}_j^{(p)}$ through minimizing the objective function in this coordinate while holding fixed all other targets' coefficients, which can be denoted as:

$$\mathbf{x}_j^{(p)} = \operatorname{argmin}_{\mathbf{x}_j^{(p)}} J(\hat{\mathbf{x}}_1^{(p)}, \ldots, \mathbf{x}_j^{(p)}, \hat{\mathbf{x}}_{j+1}^{(p-1)}, \ldots, \hat{\mathbf{x}}_M^{(p-1)}) \quad (6)$$

The corresponding objective is:

$$J(\mathbf{x}_j^{(p)}) = \frac{1}{2} \left\| \hat{\mathbf{y}} - \mathbf{D}_j' \mathbf{x}_j^{(p)} \right\|_2^2 + \lambda \left| \mathbf{x}_j^{(p)} \right|$$
$$\hat{\mathbf{y}} = \mathbf{y} - \sum_{k=1}^{j-1} \mathbf{D}_k' \hat{\mathbf{x}}_k^{(p)} - \sum_{k=j+1}^{M} \mathbf{D}_k' \hat{\mathbf{x}}_k^{(p-1)} \quad (7)$$

Through proximal gradient descent, the target coefficient $\mathbf{x}_j^{(p)}$ is updated as

$$\mathbf{x}_j^{(p)} = S_\lambda \left( \mathbf{x}_j^{(p-1)} - \mu \mathbf{D}_j'^{H} (\hat{\mathbf{y}} - \mathbf{D}_j' \mathbf{x}_j^{(p-1)}) \right)$$
$$S_\lambda(\mathbf{x}) = \operatorname{sign}(\mathbf{x}) \odot \max(\mathbf{x} - \lambda, 0) \quad (8)$$

where $S_\lambda(\mathbf{x})$ is the soft thresholding function, $\mu$ is step size, and $\odot$ denotes the Hadamard product.

## 4. EXPERIMENTS ON SIMULATION AND MEASURED DATA

In this section, experiments on both simulated and measured data are performed. The center frequency of the signal is 10 GHz, the bandwidth is 2 GHz, and the size of the scanning array is 5 m × 5 m.

### 4.1. Simulations

The imaging scene contains point targets with different positions and intensities, which covers 20 m × 20 m square and is 25 m away from the system, as shown in Fig. 3(a). This approximates the size of a jet plane. 8 scatters (−10 dBsm) located at the surrounding area, and one scatter with 4 scatters are located at the center area. It's worth noted that the it's closely adjacent to the 4 scatters. And its energy is 10 dB weaker than them. The degraded image is shown in Fig. 3(b).

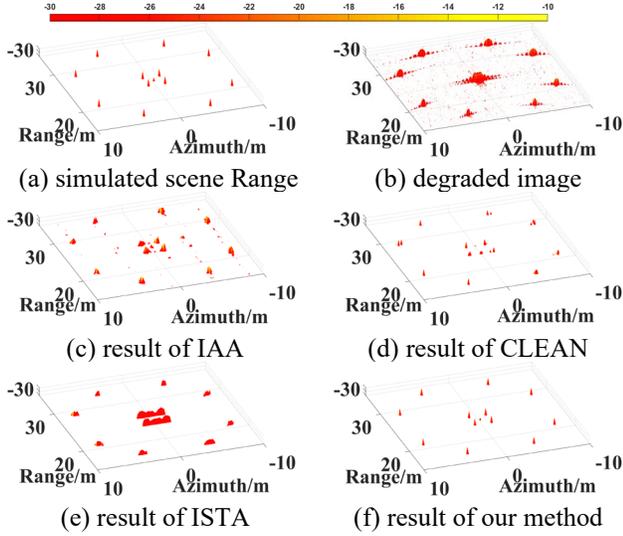

Fig 3. Simulation experiment. (a) Simulated scene; (b) degraded image; (c)-(f) results of IAA, CLEAN, ISTA and the proposed method respectively. (Calibrated amplitude in the vertical axis, unit: dB)

TABLE 1
Comparison of Mean Amplitude Error

|  | proposed | ISTA | IAA | CLEAN |
|---|---|---|---|---|
| mean amplitude error | **0.85 dB** | 1.74 dB | 2.15 dB | 4.19 dB |

As being analyzed before, the sidelobes are 2D spatial-variant, the resolution along range is also variant. What's worse, the scatter at the center is buried by sidelobes of surrounding scatters. Three different types of methods are chosen to compare with, which are ISTA [4], IAA [5] and CLEAN [6]. Results are shown in Fig. 3(c)-3(e) respectively. The result of the proposed method is shown in Fig. 3(f).

Numerous clutters remain in the result of IAA and the weak scatter is seriously under estimated. For the CLEAN, multiple fake scatters are restored while the weak scatter is not restored. For the ISTA, due to the limited resolution enhancement, closely-spaced scatters are still mixed. Still, the weak one is not restored.

Compared with others, the proposed method achieves the most superior performance. All the scatters are restored and the sidelobes and clutter are well suppressed. Besides, estimation of all the scatters' positions and amplitudes are accurate. The mean amplitude error is least (less than **0.85 dB**), as shown in Table 1. The mean position error is less than half of the wavelength.

### 4.2. Measured data experiments

The data is measured by a vehicle-borne near-field SAR system, as shown in Fig. 4. Two types of imaging scenes are considered to verify the proposed method, as Fig. 5 depicts.

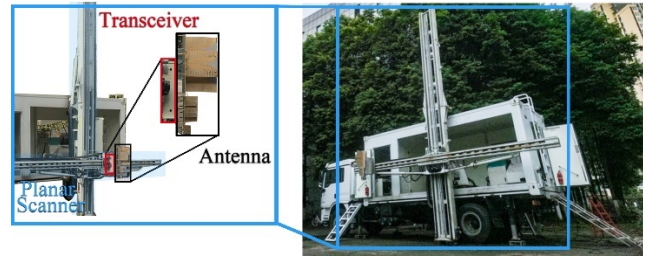

Fig.4. Vehicle-borne near-field SAR system

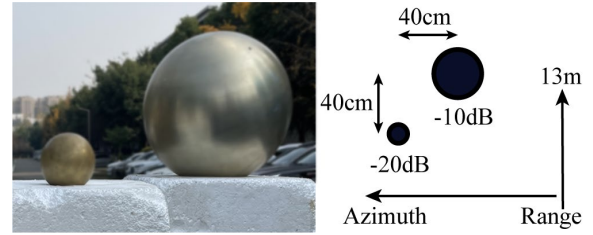

(a) scene 1

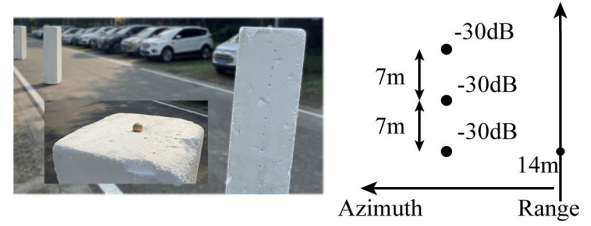

(b) scene 2

Fig.5. Imaging scenes. (a) Scene 1; (b) scene 2.

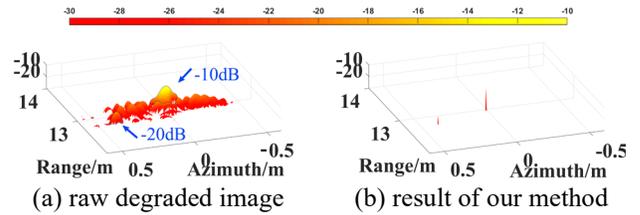

(a) raw degraded image        (b) result of our method

Fig.6. Result of imaging scene 1. (a) Raw degraded image; (b) result of the propose method. (Calibrated amplitude in the vertical axis, unit: dB)

Scene 1 [see Fig. 5(a)] contains two closely-spaced scatters of $-10$ dBsm and $-20$ dBsm. The raw degraded image is shown in Fig. 6(a). The stronger scatter's sidelobes almost bury the other one. Result of the proposed method is shown in Fig. 6(b). Both the strong scatter and the weaker scatter are restored accurately. The amplitude errors are 0.82 dB and 1.87 dB respectively. The position errors are less than 1 wavelength. Comparing with the results of the simulation experiments, limited precision losses may due to stronger clutter and noise.

The second imaging scene is shown in Fig. 5(b). Three $-30$ dBsm scatters are located at the same azimuth locations

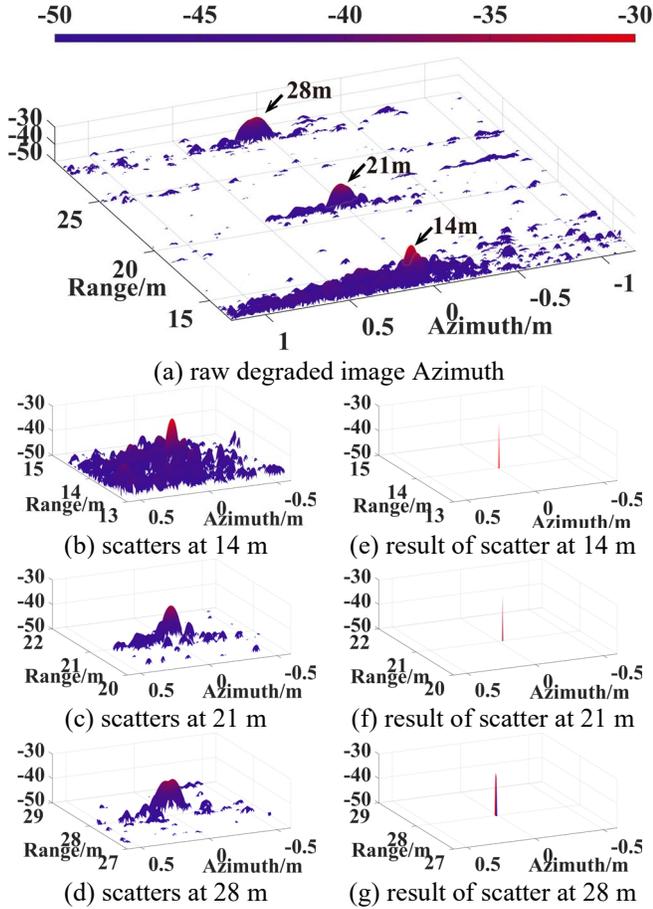

Fig 7. Result of imaging scene 2. (a) Raw degraded image; (b)-(d) enlarged views of the scatters at 14 m, 21 m and 28 m; (e)-(g) results of the scatter at 14 m, 21 m and 28 m respectively. (Calibrated amplitude in vertical axis, unit: dB)

but the different range locations of 14 m, 21 m, and 28 m respectively. Compared to the first scene, scatters' energy is less, revealing more clutter and noise. The raw degraded image is shown in Fig. 7(a)-7(d). In accordance with the analysis in section 2, the widths of range mainlobes keep unchanged, while the azimuth mainlobe widths are broadened along the range direction. The width of scatter at 28 m is approximately twice as the one at 14 m. Results of the proposed method is shown in Fig. 7(e)-(g) respectively. Clutter and noise are well suppressed that only true scatters are restored. The amplitude errors are all less than 2 dB. Specifically, their amplitude errors are 0.88 dB, 1.38 dB, 1.86 dB respectively. Besides, the position errors are all less than 0.0375 m (one wavelength).

The results on both imaging scenes verify the effectiveness and the superiority of the proposed method.

## 5. CONCLUSION

Near-field SAR has diverse practical application potential that needs higher image quality, which is the basis for interpretation. However, compared with the far-field SAR, the observation angle is larger, which brings unique characteristics and problems. To ease the current methods' limitations, we propose an image restoration method from the perspective of deconvolution. The unique characteristic of 2D spatial variations is analytically formed as a complex convolution process of the 2D spatial-variant degradation operations. To restore targets, deconvolution is implemented through cyclic coordinate descend algorithms with targets' sparsity into consideration. Simulations and experiments on measured data verify the effectiveness and the superiority of the proposed method, compared with current methods. With our method, large-scale targets can be restored well. Also, targets of different energies that are closely spaced can also be restored well. These reveal the potential of our method for assisting the more incisive comprehension of near-field SAR images.

## 6. REFERENCES

[1] Z. Li, J. Wang, J. Wu, and Q. H. Liu, "A fast radial scanned near-field 3-D SAR imaging system and the reconstruction method," *IEEE Trans. Geosci. Remote Sens.*, vol. 53, no. 3, pp. 1355–1363, 2015.

[2] B. Ding and G. Wen, "Target reconstruction based on 3-D scattering center model for robust SAR ATR," *IEEE Trans. Geosci. Remote Sens.*, vol. 56, no. 7, pp. 3772–3785, 2018.

[3] M. Wang et al., "TPSSI-Net: Fast and enhanced two-path iterative network for 3D SAR sparse imaging," *IEEE Trans. Image Process.*, vol. 30, pp. 7317–7332, 2021.

[4] Y. Wang et al., "An RCS measurement method using sparse imaging based 3-D SAR complex image," *IEEE Antennas Wirel. Propag. Lett.*, vol. 21, no. 1, pp. 24–28, 2022.

[5] W. Roberts, P. Stoica, J. Li, T. Yardibi, and F. A. Sadjadi, "Iterative adaptive approaches to MIMO radar imaging," *IEEE J. Sel. Top. Signal Process.*, vol. 4, no. 1, pp. 5–20, 2010.

[6] D.-J. Yun, J.-I. Lee, K.-U. Bae, J.-H. Yoo, K.-I. Kwon, and N.-H. Myung, "Improvement in computation time of 3-D scattering center extraction using the shooting and bouncing ray technique," *IEEE Trans. Antennas Propag.*, vol. 65, no. 8, pp. 4191–4199, 2017.

[7] H. Yang, C. Chen, S. Chen, F. Xi, and Z. Liu, "Non-Common Band SAR Interferometry Via Compressive Sensing," *IEEE Trans. Geosci. Remote Sens.*, vol. 58, no. 6, pp. 4436–4453, 2020.

[8] Y. Wang, Z. He, X. Zhan, Y. Fu, and L. Zhou, "Three-dimensional sparse SAR imaging with generalized Lq regularization," *Remote Sens. (Basel)*, vol. 14, no. 2, p. 288, 2022.

[9] T. Hastie, R. Tibshirani, and M. Wainwright, *Statistical learning with sparsity: The lasso and generalizations*. London, England: CRC Press, 2020.